\documentclass[12pt,a4paper]{article}
\usepackage{url}
\makeindex
\newcommand{\ahref}[2]{{#2}}%
\newcommand{\aname}[2]{{#2}}%
\newcommand{\ahrefurl}[1]{\url{#1}}%
\newcommand{\mailto}[1]{\texttt{#1}}%
\newcommand{\imgsrc}[2][]{}%
\newcommand{\home}[1]{\protect\raisebox{-.75ex}{\char126}#1}
\providecommand{\preprintno}[1]{\relax}
\usepackage{verbatim,array,amsmath,amssymb,graphicx}
\usepackage{thophys}
\usepackage{feynmp,emp}
\empaddtoTeX{\usepackage{amsmath,amssymb}}
\empaddtoTeX{\usepackage{thophys}}
\empaddtoprelude{input graph;}
\empaddtoprelude{input boxes;}
\makeatletter
  \@ifundefined{frontmatter}%
    {%
     }
    {}
\makeatother
\allowdisplaybreaks

\newenvironment{algorithm}[1]%
 {\begin{list}{}%
   {\setlength{\leftmargin}{3em}%
    \setlength{\rightmargin}{3em}%
    \setlength{\itemindent}{1em}%
    \setlength{\listparindent}{0pt}%
    \settowidth{\labelwidth}{5em}%
    }}%
 {\end{list}}
\newenvironment{files}%
 {\begin{list}{}%
   {\setlength{\leftmargin}{3em}%
    \setlength{\rightmargin}{3em}%
    \setlength{\itemindent}{1em}%
    \setlength{\listparindent}{0pt}%
    \settowidth{\labelwidth}{5em}%
    }}%
 {\end{list}}
\newenvironment{options}%
 {\begin{list}{}%
   {\setlength{\leftmargin}{3em}%
    \setlength{\rightmargin}{3em}%
    \setlength{\itemindent}{1em}%
    \setlength{\listparindent}{0pt}%
    \settowidth{\labelwidth}{5em}%
    }}%
 {\end{list}}
\newenvironment{code}{\verbatim}{\endverbatim\noindent}

\newcommand{\eprint}[1]{\ahref{http://arXiv.org/abs/#1}{#1}}
\begin{document}
%BEGIN IMAGE
\begin{fmffile}{previewpics}
\fmfset{arrow_ang}{10}
\fmfset{curly_len}{2mm}
\fmfset{wiggly_len}{3mm}
\begin{empfile}
%END IMAGE
%%%%%%%%%%%%%%%%%%%%%%%%%%%%%%%%%%%%%%%%%%%%%%%%%%%%%%%%%%%%%%%%%%%%%%%%
\title{O'Mega:\\An Optimizing Matrix Element Generator}
\author{%
  Mauro Moretti\thanks{e-mail: \mailto{moretti@fe.infn.it}}\\
  \hfil\\
    Dipartimento di Fisica, Universit\`a di Ferrara\\
    and INFN, Sezione di Ferrara, Ferrara, Italy\\
  \hfil\\\hfil\\
  \ahref{http://heplix.ikp.physik.tu-darmstadt.de/\home{ohl}/}%
    {Thorsten Ohl}\thanks{e-mail: \mailto{ohl@hep.tu-darmstadt.de}}\\
  J\"urgen Reuter\thanks{e-mail: \mailto{reuter@hep.tu-darmstadt.de}}\\
  \hfil\\
    \ahref{http://www.tu-darmstadt.de}{Darmstadt University of Technology}\\
    Schlo\ss{}gartenstr. 9, D-64289 Darmstadt, Germany}
\preprintno{\hfil}
\date{\fbox{IKDA 2001/06\quad{} LC-TOOL-2001-040\quad{} \eprint{hep-ph/0102195}}}
\maketitle
\begin{abstract}
  We sketch the architecture of \textit{O'Mega}, a new
  optimizing compiler for tree amplitudes in quantum field theory,
  and briefly describe its usage.
  O'Mega generates the most efficient code currently available for
  scattering amplitudes for many polarized particles in the Standard
  Model and its extensions.
\end{abstract}
%%%%%%%%%%%%%%%%%%%%%%%%%%%%%%%%%%%%%%%%%%%%%%%%%%%%%%%%%%%%%%%%%%%%%%%%

%HEVEA O'Mega is Free Software and the
%HEVEA \ahref{ftp://heplix.ikp.physik.tu-darmstadt.de/pub/ohl/omega}{sources}
%HEVEA can be found at
%HEVEA \ahref{ftp://heplix.ikp.physik.tu-darmstadt.de/pub/ohl/omega}{this link}.
%HEVEA Follow \ahrefloc{installation}{this link} for
%HEVEA \ahrefloc{installation}{installation instructions}.

%%%%%%%%%%%%%%%%%%%%%%%%%%%%%%%%%%%%%%%%%%%%%%%%%%%%%%%%%%%%%%%%%%%%%%%%
%%%\tableofcontents
\newpage
\section{Introduction}
\label{sec:intro}
Current and planned experiments in high energy physics can probe
physics in
processes with polarized beams and many tagged particles in the final
state.  The combinatorial explosion of the number of Feynman diagrams
contributing to scattering amplitudes for many external particles
calls for the development of more compact representations that
translate well to efficient and reliable numerical code.  In gauge
theories, the contributions from individual Feynman diagrams are gauge
dependent.  Strong numerical cancellations in a redundant
representation built from individual Feynman diagrams lead to a loss
of numerical precision, stressing further the need for eliminating
redundancies.

Due to the large number of processes that have to be studied in order
to unleash the potential of modern experiments, the construction of
nearly optimal representations must be possible algorithmically on a
computer and should not require human ingenuity for each new
application.

\textit{O'Mega}~\cite{O'Mega,Ohl:2000:ACAT,Ohl:2000:LCWS} is a compiler for
tree-level scattering amplitudes that satisfies these requirements.
O'Mega is independent of the target language and can therefore create
code in any programming language for which a simple output module has
been written.  To support a physics model, O'Mega requires as input
only the Feynman rules and the relations among coupling constants.

Similar to the earlier numerical approaches~\cite{ALPHA:1997}
and~\cite{HELAC:2000}, O'Mega reduces the growth in calculational
effort from a factorial of the number of particles to an exponential.
The symbolic nature of O'Mega, however, increases its flexibility.
Indeed, O'Mega can emulate both~\cite{ALPHA:1997}
and~\cite{HELAC:2000} and produces code that is empirically at least
twice as fast.  The detailed description of all algorithms is
contained in the extensively commented source code of
O'Mega~\cite{O'Mega}.

In this note, we sketch the architecture of O'Mega and describe the
usage of the first version.  The building blocks of the representation
of scattering amplitudes generated by O'Mega are described in
section~\ref{sec:1POW} and directed acyclical graphs are introduced in
section~\ref{sec:DAG}.  The algorithm for constructing the directed
acyclical graph is presented in section~\ref{sec:algorithm} and its
implementation is described in section~\ref{sec:implementation}.
We conclude with a few results and examples in
section~\ref{sec:results}.  Practical information is
presented in the appendices: installation of the O'Mega software in
appendix~\ref{sec:installation}, running of the O'Mega compiler in
appendix~\ref{sec:running} and using O'Mega's output in
appendix~\ref{sec:using}.  Finally, appendix~\ref{sec:extensions}
briefly discusses mechanisms for extending O'Mega.

%%%%%%%%%%%%%%%%%%%%%%%%%%%%%%%%%%%%%%%%%%%%%%%%%%%%%%%%%%%%%%%%%%%%%%%%
\section{One Particle Off Shell Wave Functions}
\label{sec:1POW}

\textit{One Particle Off-Shell Wave Functions}~(1POWs) are obtained
from connected Greensfunctions by applying the LSZ reduction formula
to all but one external line while the remaining line is kept off the
mass shell
\begin{multline}
  W(x; p_1,\ldots,p_n; q_1,\ldots,q_m) = \\
    \Braket{\phi(q_1),\ldots,\phi(q_m);\text{out}|\Phi(x)
           |\phi(p_1),\ldots,\phi(p_n);\text{in}}\,.
\end{multline}
Depending on the context, the off shell line will either be understood as
amputated or not.  For example,
$\Braket{\phi(q_1),\phi(q_2);\text{out}|\Phi(x)|\phi(p_1);\text{in}}$
in unflavored scalar $\phi^3$-theory is given at tree level by
%HEVEA\begin{center}
%BEGIN IMAGE
\begin{equation}
  \parbox{26\unitlength}{%
    \fmfframe(2,4)(6,5){%
      \begin{fmfgraph*}(17,15)
       \fmflabel{$x$}{x}
       \fmflabel{$p_1$}{l}
       \fmflabel{$q_1$}{r1}
       \fmflabel{$q_2$}{r2}
       \fmftop{x}
       \fmfleft{l,dl}
       \fmfright{r1,r2,dr}
       \fmf{plain}{l,v}
       \fmf{plain}{r1,v}
       \fmf{plain}{r2,v}
       \fmf{plain,tension=3}{x,v}
       \fmfblob{.4w}{v}
       \fmfdot{x}
      \end{fmfgraph*}}} =
  \parbox{26\unitlength}{%
    \fmfframe(2,4)(6,5){%
      \begin{fmfgraph*}(17,15)
       \fmflabel{$x$}{x}
       \fmflabel{$p_1$}{l}
       \fmflabel{$q_1$}{r1}
       \fmflabel{$q_2$}{r2}
       \fmftop{x}
       \fmfleft{l,dl}
       \fmfright{r1,r2,dr}
       \fmf{plain}{l,v}
       \fmf{plain}{r1,vr,v}
       \fmf{plain}{r2,vr}
       \fmf{plain,tension=5}{x,v}
       \fmfdot{x}
      \end{fmfgraph*}}} +
  \parbox{26\unitlength}{%
    \fmfframe(2,4)(6,5){%
      \begin{fmfgraph*}(17,15)
       \fmflabel{$x$}{x}
       \fmflabel{$p_1$}{l}
       \fmflabel{$q_1$}{r1}
       \fmflabel{$q_2$}{r2}
       \fmftop{x}
       \fmfleft{l,dl}
       \fmfright{r1,r2,dr}
       \fmf{plain}{l,vr,v}
       \fmf{plain}{r1,vr}
       \fmf{plain}{r2,v}
       \fmf{plain,tension=5}{x,v}
       \fmfdot{x}
      \end{fmfgraph*}}} +
  \parbox{26\unitlength}{%
    \fmfframe(2,4)(6,5){%
      \begin{fmfgraph*}(17,15)
       \fmflabel{$x$}{x}
       \fmflabel{$p_1$}{l}
       \fmflabel{$q_1$}{r1}
       \fmflabel{$q_2$}{r2}
       \fmftop{x}
       \fmfleft{l,dl}
       \fmfright{r1,r2,dr}
       \fmf{plain}{l,vr}
       \fmf{plain,tension=0.5}{vr,v}
       \fmf{plain}{r1,v}
       \fmf{plain,rubout,tension=0.5}{r2,vr}
       \fmf{plain,tension=5}{x,v}
       \fmfdot{x}
      \end{fmfgraph*}}}.
\end{equation}
%END IMAGE
%HEVEA\imageflush
%HEVEA\end{center}

The number of distinct momenta that can be formed from
$n$~external momenta is $P(n)=2^{n-1}-1$.  Therefore, the number of
tree 1POWs grows exponentially with the number of external particles
and not with a factorial, as the number of Feynman diagrams, e.\,g.{}
$F(n)=(2n-5)!!=(2n-5)\cdot\ldots5\cdot3\cdot1$ in unflavored
$\phi^3$-theory.

At tree-level, the set of all 1POWs for a given set of external
momenta can be constructed recursively
%HEVEA\begin{center}
%BEGIN IMAGE
\begin{equation}
\label{eq:recursive-1POW}
  \parbox{22\unitlength}{%
    \fmfframe(2,3)(2,1){%
      \begin{fmfgraph*}(17,15)
       \fmflabel{$x$}{x}
       \fmftop{x}
       \fmfbottomn{n}{6}
       \fmf{plain,tension=6}{x,n}
       \fmfv{d.sh=circle,d.f=empty,d.si=30pt,l=$n$,l.d=0}{n}
       \begin{fmffor}{i}{1}{1}{6}
         \fmf{plain}{n,n[i]}
       \end{fmffor}
      \end{fmfgraph*}}} = 
  \sum_{k+l=n}
  \parbox{32\unitlength}{%
    \fmfframe(2,3)(2,1){%
      \begin{fmfgraph*}(27,15)
       \fmflabel{$x$}{x}
       \fmftop{x}
       \fmfbottomn{n}{6}
       \fmf{plain,tension=8}{x,n}
       \fmf{plain,tension=4}{n,k}
       \fmf{plain,tension=4}{n,l}
       \fmfv{d.sh=circle,d.f=empty,d.si=20pt,l=$k$,l.d=0}{k}
       \fmfv{d.sh=circle,d.f=empty,d.si=20pt,l=$l$,l.d=0}{l}
       \fmffixed{(30pt,0pt)}{k,l}
       \begin{fmffor}{i}{1}{1}{4}
         \fmf{plain}{k,n[i]}
       \end{fmffor}
       \begin{fmffor}{i}{5}{1}{6}
         \fmf{plain}{l,n[i]}
       \end{fmffor}
       \fmfdot{n}
      \end{fmfgraph*}}}\,,
\end{equation}
%END IMAGE
%HEVEA\imageflush
%HEVEA\end{center}
where the sum extends over all partitions of the set of $n$~momenta.
This recursion will terminate at the external wave functions.

For all quantum field theories, there are---well defined, but not
unique---sets of \emph{Keystones}~$K$~\cite{O'Mega} such that the sum
of tree Feynman diagrams for a given process can be expressed as a
sparse sum of products of 1POWs without double counting.  In a theory
with only cubic couplings this is expressed as
\begin{equation}
\label{eq:keystones}
  T = \sum_{i=1}^{F(n)} D_i =
      \sum_{k,l,m=1}^{P(n)}
        K^{3}_{f_kf_lf_m}(p_k,p_l,p_m)
        W_{f_k}(p_k)W_{f_l}(p_l)W_{f_m}(p_m)\,,
\end{equation}
with obvious generalizations.
The non-trivial problem is to avoide the
double counting of diagrams like
%HEVEA\begin{center}
%BEGIN IMAGE
\begin{center}
   \begin{fmfgraph}(25,16)
     \fmfleftn{l}{3}
     \fmfrightn{r}{3}
     \fmf{plain}{l1,v4}
     \fmf{plain}{l2,v4}
     \fmf{plain}{l3,v4}
     \fmf{plain}{r1,v1}
     \fmf{plain}{r2,v1}
     \fmf{plain}{v1,v2}
     \fmf{plain}{r3,v2}
     \fmf{plain}{v2,v4}
     \fmfv{d.sh=circle,d.fill=empty,d.si=6thin}{v4}  
     \fmfdot{v1,v2}
   \end{fmfgraph}
   \qquad\qquad
   \begin{fmfgraph}(25,16)
     \fmfleftn{l}{3}
     \fmfrightn{r}{3}
     \fmf{plain}{l1,v4}
     \fmf{plain}{l2,v4}
     \fmf{plain}{l3,v4}
     \fmf{plain}{r1,v1}
     \fmf{plain}{r2,v1}
     \fmf{plain}{v1,v2}
     \fmf{plain}{r3,v2}
     \fmf{plain}{v2,v4}
     \fmfv{d.sh=circle,d.fill=empty,d.si=6thin}{v2}  
     \fmfdot{v1,v4}
   \end{fmfgraph}\,,
\end{center} 
%END IMAGE
%HEVEA\imageflush
%HEVEA\end{center}
where the circle denotes the keystone. The problem has been solved
explicitely for general theories with vertices of arbitrary
degrees~\cite{O'Mega}.  The solution is inspired by
arguments~\cite{ALPHA:1997} based on the equations of motion (EOM) of
the theory in the presence of sources. The iterative solution of the
EOM leads to the construcion of the 1POWs and the constraints imposed
on the 1POWs by the EOM suggest the correct set~\cite{ALPHA:1997} of
partitions $\{(p_k,p_l,p_m)\}$ in equation~(\ref{eq:keystones}).

The maximally symmetric solution selects among equivalent diagrams the
keystone closest to the center of a diagram.  This corresponds to
the numerical expressions of~\cite{ALPHA:1997}.  The absence of double
counting can be demonstrated by counting the number~$F(d_{\max},n)$ of
unflavored Feynman tree diagrams with~$n$ external legs and vertices of
maximum degree~$d_{\max}$ in to different ways: once directly and then
as a sum over keystones.  The number~$\tilde F(d_{\max},N_{d,n})$ of
unflavored Feynman tree diagrams for one keystone
$N_{d,n}=\{n_1,n_2,\ldots,n_d\}$, with $n = n_1 + n_2 + \cdots + n_d$,
is given by the product of the number of subtrees and symmetry factors
\begin{subequations}
\begin{equation}
  \tilde F(d_{\max},N_{d,n}) =
    \frac{n!}{|\mathcal{S}(N_{d,n})|\sigma(n_d,n)}
    \prod_{i=1}^{d} \frac{F(d_{\max},n_i+1)}{n_i!}\,
\end{equation}
where $|\mathcal{S}(N)|$ is the size of the symmetric group
of~$N$, $\sigma(n,2n) = 2$ and $\sigma(n,m) = 1$ otherwise.  Indeed,
it can be verified that the sum over all keystones reproduces the
number
\begin{equation}
  F(d_{\max},n) =
    \sum_{d=3}^{d_{\max}}
    \sum_{\substack{N = \{n_1,n_2,\ldots,n_d\}\\
                    n_1 + n_2 + \cdots + n_d = n\\
                    1 \le n_1 \le n_2 \le \cdots \le n_d \le \lfloor n/2 \rfloor}}
     \tilde F(d_{\max},N)
\end{equation}
\end{subequations}
of \emph{all} unflavored Feynman tree diagrams.

A second consistent prescription for the construction of keystones is
maximally asymmetric and selects the keystone adjacent to a chosen
external line.  This prescription reproduces the approach
in~\cite{HELAC:2000} where the tree-level Schwinger-Dyson equations
are used as a special case of the EOM.

Recursive algorithms for gauge theory amplitudes have been pioneered
in~\cite{Berends:1988me}.  The use of 1POWs as basic building blocks
for the calculation of scattering amplitudes in tree approximation has
been advocated in~\cite{HELAS} and a heuristic procedure, without
reference to keystones, for minimizing the number of arithmetical
operations has been suggested.  This approach is used by
MADGRAPH~\cite{MADGRAPH:1994} for fully automated calculations. The
heuristic optimizations are quite efficient for $2\to4$ processes, but
the number of operations remains bounded from below by the number of
Feynman diagrams.

%%%%%%%%%%%%%%%%%%%%%%%%%%%%%%%%%%%%%%%%%%%%%%%%%%%%%%%%%%%%%%%%%%%%%%%%
\subsection{Ward Identities}
\label{sec:WI}

\begin{subequations}
A particularly convenient property of the 1POWs in gauge theories is
that, even for vector particles, the 1POWs are `almost' physical
objects and satisfy simple Ward Identities
\label{eq:ward}
\begin{equation}
    \frac{\partial}{\partial x_\mu}
    \Braket{\text{out}|A_\mu(x)|\text{in}}_{\text{amp.}} = 0
\end{equation}
for unbroken gauge theories and
\begin{equation}
    \frac{\partial}{\partial x_\mu}
    \Braket{\text{out}|W_\mu(x)|\text{in}}_{\text{amp.}} =
      - m_W \Braket{\text{out}|\phi_W(x)|\text{in}}_{\text{amp.}}
\end{equation}
for spontaneously broken gauge theories in $R_\xi$-gauge for all
physical external states~$\ket{in}$ and $\ket{out}$. Thus the
identities~(\ref{eq:ward}) can serve as powerful numerical checks
both for the consistency of a set of Feynman rules and for the
numerical stability of the generated code.   The code for matrix
elements can optionally be instrumented by O'Mega with numerical
checks of these Ward identities for intermediate lines.
\end{subequations}

%%%%%%%%%%%%%%%%%%%%%%%%%%%%%%%%%%%%%%%%%%%%%%%%%%%%%%%%%%%%%%%%%%%%%%%%
\section{Directed Acyclical Graphs}
\label{sec:DAG}

The algebraic expression for the tree-level scattering amplitude in
terms of Feynman diagrams is itself a tree.  The much slower growth of
the set of 1POWs compared to the set of Feynman diagrams shows that this
representation is extremely redundant. In this case, \emph{Directed
Acyclical Graphs} (DAGs) provide a more efficient representation, as
illustrated by a trivial example
%HEVEA\begin{center}
%BEGIN IMAGE
\begin{empcmds}
  vardef dag_coords =
    pair node[][]; node[1][1] = (.5w,.h);
    node[2][1] = (.3w,2/3h); node[2][2] = (.7w,2/3h);
    node[3][1] = (.2w,1/3h); node[3][2] = (.4w,1/3h);
    node[3][3] = (.6w,1/3h); node[3][4] = (.8w,1/3h);
    node[4][1] = (.5w,0/3h); node[4][2] = (.7w,0/3h);
    % setbounds currentpicture to (0,0)--(w,0)--(w,h)--(0,h)--cycle;
  enddef;
  vardef dag_common =
    dag_coords;
    pickup pencircle scaled 1pt;
    label.rt (btex $\times$ etex, node[1][1]);
    draw node[1][1]--node[2][2];
    label.rt (btex $+$ etex, node[2][2]);
    draw node[2][2]--node[3][3];
    draw node[2][2]--node[3][4];
    label.rt (btex $\times$ etex, node[3][3]);
    draw node[3][3]--node[4][1];
    draw node[3][3]--node[4][2];
    label.rt (btex $\vphantom{b}c$ etex, node[3][4]);
    label.rt (btex $\vphantom{b}a$ etex, node[4][1]);
    label.rt (btex $\vphantom{b}b$ etex, node[4][2]);
    pickup pencircle scaled 3pt;
    pickup pencircle scaled 3pt;
    drawdot node[1][1];
    drawdot node[2][2];
    drawdot node[3][3];
  enddef;
\end{empcmds}
\begin{empdef}[dag](38,16)
  dag_common;
  pickup pencircle scaled 1pt;
  draw node[1][1]{(-1,-1)}..{(1,-1)}node[3][3];
\end{empdef}
\begin{empdef}[tree](38,16)
  dag_common;
  pickup pencircle scaled 1pt;
  label.rt (btex $\times$ etex, node[2][1]);
  draw node[1][1]--node[2][1];
  draw node[2][1]--node[3][1];
  draw node[2][1]--node[3][2];
  label.rt (btex $\vphantom{b}a$ etex, node[3][1]);
  label.rt (btex $\vphantom{b}b$ etex, node[3][2]);
  pickup pencircle scaled 3pt;
  drawdot node[2][1];
\end{empdef}
\begin{equation}
  ab (ab+c) =
  \parbox{28\unitlength}{\hfil\empuse{tree}\hfil}
    = \parbox{18\unitlength}{\hfil\empuse{dag}\hfil}
\end{equation}
%END IMAGE
%HEVEA\imageflush
%HEVEA\end{center}
where one multiplication is saved.  The replacement of expression
trees by equivalent DAGs is part of the repertoire of optimizing
compilers, known as \emph{common subexpression elimination}.
Unfortunately, this approach fails in practice for all interesting
expressions appearing in quantum field theory, because of the
combinatorial growth of space and time required to find an almost
optimal factorization.

However, the recursive definition in equation~(\ref{eq:recursive-1POW})
allows to construct the DAG of the 1POWs in equation~(\ref{eq:keystones})
\emph{directly}~\cite{O'Mega}, without having to construct and
factorize the Feynman diagrams explicitely.

As mentioned above, there is more than one consistent prescription for
constructing the set of keystones~\cite{O'Mega}.  The symbolic
expressions constructed by O'Mega contain the symbolic equivalents of
the numerical expressions computed by~\cite{ALPHA:1997} (maximally
symmetric keystones) and~\cite{HELAC:2000} (maximally asymmetric
keystones) as special cases.

%%%%%%%%%%%%%%%%%%%%%%%%%%%%%%%%%%%%%%%%%%%%%%%%%%%%%%%%%%%%%%%%%%%%%%%%
\section{Algorithm}
\label{sec:algorithm}

By virtue of their recursive construction in
Eqs.~(\ref{eq:recursive-1POW}), tree-level 1POWs form a DAG and the
problem is to find the smallest DAG that corresponds to a given tree,
(i.\,e.~a given sum of Feynman diagrams).  O'Mega's algorithm
proceeds in four steps
\begin{algorithm}{Calculate}
  \item[Grow] starting from the external particles, build the tower of
    \emph{all} 1POWs up to a given height (the height
    is less than the number of external lines for asymmetric
    keystones and less than half of that for symmetric keystones)
    and translate it to the equivalent DAG~$D$.
  \item[Select] from $D$, determine \emph{all} possible
    \emph{flavored keystones} for the process under
    consideration and the 1POWs appearing in them.
  \item[Harvest] construct a sub-DAG $D^*\subseteq D$ consisting
    \emph{only} of nodes that contribute to the 1POWs
    appearing in the flavored keystones.
  \item[Calculate] multiply the 1POWs as specified by the keystones
    and sum the keystones.
\end{algorithm}
By construction, the resulting expression contains no more
redundancies and can be translated to a numerical expression.  In
general, asymmetric keystones create an expression that is smaller
by a few percent than the result from symmetric keystones, but it
is not yet clear which approach produces the numerically more robust
results.

The details of this algorithm as implemented in O'Mega are described
in the source code~\cite{O'Mega}.  The persistent data
structures~\cite{Okasaki:1998:book} used for the determination
of~$D^*$ are very efficient so that the generation of, e.\,g.~Fortran
code for amplitudes in the Standard Model is always much faster than
the subsequent compilation.

%%%%%%%%%%%%%%%%%%%%%%%%%%%%%%%%%%%%%%%%%%%%%%%%%%%%%%%%%%%%%%%%%%%%%%%%
\section{Implementation}
\label{sec:implementation}
The O'Mega compiler is implemented in O'Caml~\cite{O'Caml}, a
functional programming language of the ML family with a very
efficient, portable and freely available implementation, that can be
bootstrapped on all modern computers in a few minutes.
The library modules built on experience
from~\cite{Ohl:LOTR,Ohl:bocages}.

The powerful module system of O'Caml allows an efficient and concise
implementation of the DAGs for a specific physics model as a functor
application~\cite{O'Mega}.  This functor maps from the category of
trees to the category of DAGs and is applied to the set of trees
defined by the Feynman rules of any model under consideration.

\begin{figure}
%BEGIN IMAGE
  \includegraphics[width=\textwidth]{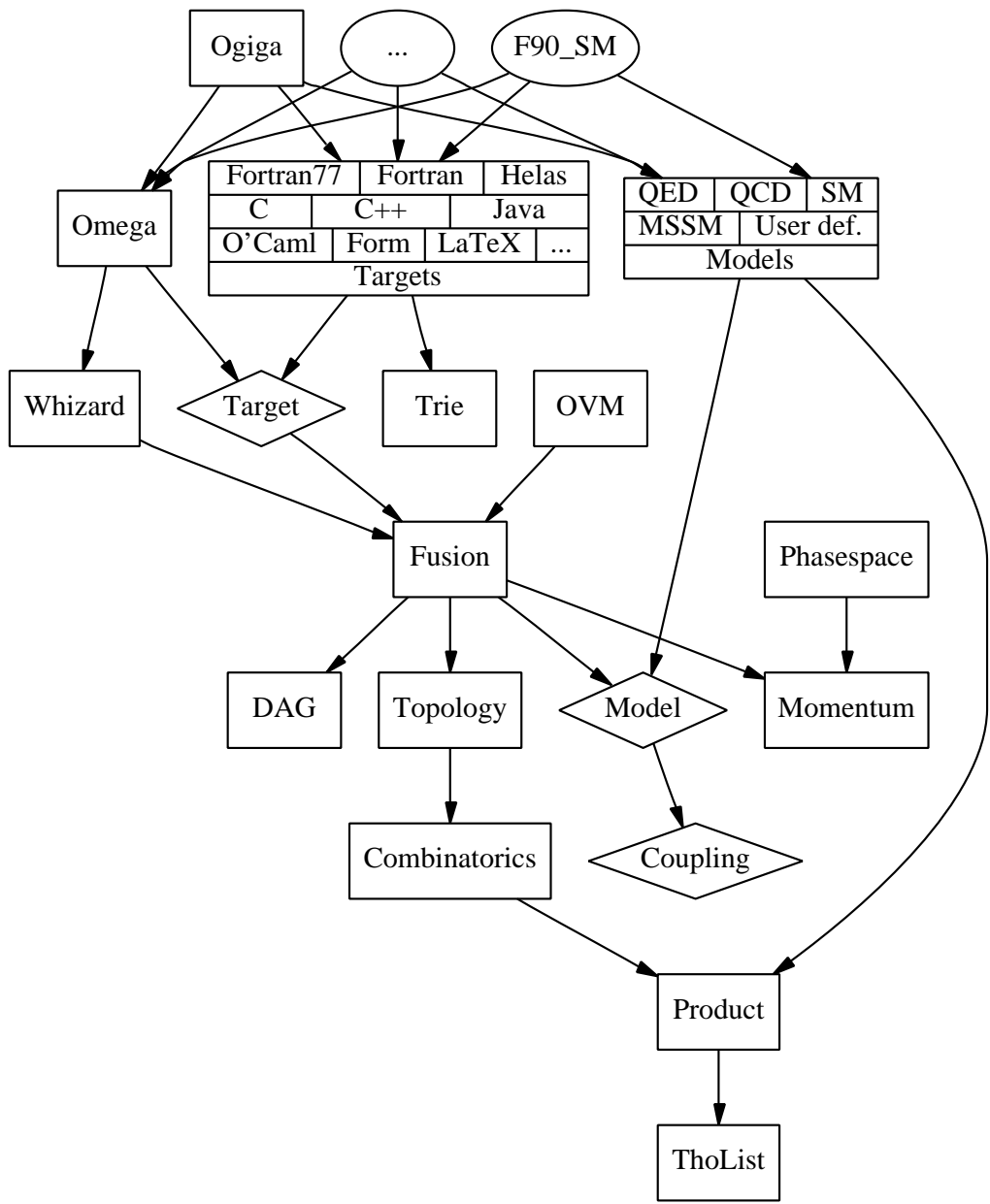}
  %includegraphics[height=.9\textheight]{modules}
%END IMAGE
%HEVEA\imageflush
  \caption{\label{fig:modules}%
    Module dependencies in O'Mega.  The diamond shaped nodes denote
    abstract signatures defining functor domains and co-domains.
    The rectangular boxes denote modules and functors, while oval
    boxes stand for example applications.}
\end{figure}
The module system of O'Caml has been used to make the combinatorial
core of O'Mega demonstrably independent from the specifics of both the
physics model and the target language~\cite{O'Mega}, as shown in
Figure~\ref{fig:modules}.  A Fortran90/95 backend has been realized
first, backends for C++ and Java will follow.  The complete
electroweak Standard Model has been implemented together with
anomalous gauge boson couplings.  The implementation of interfering
color amplitudes is currently being completed.

Many extensions of the Standard Model, most prominently the Minimal
Supersymmetric Standard Model~(MSSM), contain Majorana fermions.  In
this case, fermion lines have no canonical orientation and the
determination of the relative signs of interfering amplitudes is not
trivial.  However, the Feynman rules for Majorana fermions and fermion
number violating interactions proposed in~\cite{Denner/etal:Majorana}
have been implemented in O'Mega in analogy to the naive Feynman rules
for Dirac fermions and both methods are available.  Numerical
comparisons of amplitudes for Dirac fermions calculated both ways show
agreement at a small multiple of the machine precision.  Thus, all
ingredients for the MSSM are available in O'Mega and the
implementation of the complete MSSM Lagrangian is currently under way.
Non-minimal gauge models, including left-right symmetric models, can
be implemented easily.

As mentioned above, the compilers for the target programming language
are the slowest step in the generation of executable code.  On the
other hand, the execution speed of the code is limited by non-trivial
vertex evaluations for vectors and spinors, which need $O(10)$ complex
multiplications.  Therefore, an \emph{O'Mega Virtual Machine} can
challenge native code and avoid compilations.

%%%%%%%%%%%%%%%%%%%%%%%%%%%%%%%%%%%%%%%%%%%%%%%%%%%%%%%%%%%%%%%%%%%%%%%%
\section{Results}
\label{sec:results}

%%%%%%%%%%%%%%%%%%%%%%%%%%%%%%%%%%%%%%%%%%%%%%%%%%%%%%%%%%%%%%%%%%%%%%%%
\subsection{Examples}
\label{sec:examples}

\begin{table}
  \begin{center}
    \begin{tabular}{l|rr|rr}
                    \multicolumn{1}{c|}{process}
                  & \multicolumn{2}{c|}{Diagrams}
                  & \multicolumn{2}{c}{O'Mega} \\
                  & \multicolumn{1}{c}{\#} & vertices
                  & \#prop. & vertices \\%\hline
      $e^+e^-\to e^+\bar\nu_e d\bar u$
        &     20 &      80 &  14 &   45 \\
%%%SM4  &     20 &      80 &  14 &   45 \\
%%%SM4h &     20 &      80 &  14 &   35 \\
      $e^+e^-\to e^+\bar\nu_e d\bar u \gamma$
        &    146 &     730 &  36 &  157 \\
%%%SM4  &    142 &     710 &  33 &  151 \\
%%%SM4h &    142 &     710 &  33 &  115 \\
      $e^+e^-\to e^+\bar\nu_e d\bar u \gamma\gamma$
        &   1256 &    7536 &  80 &  462 \\
%%%SM4  &   1174 &    7044 &  71 &  441 \\
%%%SM4h &   1174 &    7044 &  71 &  361 \\
      $e^+e^-\to e^+\bar\nu_e d\bar u \gamma\gamma\gamma$
        &  12420 &   86940 & 168 & 1343 \\
%%%SM4  &  11058 &   77406 & 147 & 1284 \\
%%%SM4h &  11058 &   77406 & 147 & 1106 \\
      $e^+e^-\to e^+\bar\nu_e d\bar u \gamma\gamma\gamma\gamma$
        & 138816 & 1110528 & 344 & 3933
    \end{tabular}
  \end{center}
  \caption{\label{tab:4fgamma}%
    Radiative corrections to four fermion production: comparison of
    the computational complexity of scattering amplitudes obtained
    from Feynman diagrams and from O'Mega. (The counts correspond to
    the full Standard Model---sans light fermion Yukawa couplings---in
    unitarity gauge with quartic couplings emulated by cubic 
    couplings of non-propagating auxiliary fields.)}
\end{table}

\begin{table}
  \begin{center}
    \begin{tabular}{l|rr|rr}
                    \multicolumn{1}{c|}{process}
                  & \multicolumn{2}{c|}{Diagrams}
                  & \multicolumn{2}{c}{O'Mega} \\
                  & \multicolumn{1}{c}{\#} & vertices
                  & \#prop. & vertices \\%\hline
      $e^+e^-\to e^+\bar\nu_e d\bar u b\bar b$
        &    472 &   2832 &  49 &   232 \\
%%%SM4  &    464 &   2784 &  46 &   227 \\
%%%SM4h &    464 &   2784 &  46 &   186 \\
      $e^+e^-\to e^+\bar\nu_e d\bar u b\bar b \gamma$
        &   4956 &  34692 & 108 &   722 \\
%%%SM4  &   4738 &  33166 &  99 &   709 \\
%%%SM4h &   4738 &  33166 &  99 &   606 \\
      $e^+e^-\to e^+\bar\nu_e d\bar u b\bar b \gamma\gamma$
        &  58340 & 466720 & 226 &  2212
    \end{tabular}
  \end{center}
  \caption{\label{tab:6fgamma}%
    Radiative corrections to six fermion production: comparison of
    the computational complexity of scattering amplitudes obtained
    from Feynman diagrams and from O'Mega. (The counts correspond to
    the full Standard Model---sans light fermion Yukawa couplings---in
    unitarity gauge with quartic couplings emulated by cubic 
    couplings of non-propagating auxiliary fields.)}
\end{table}

Tables~\ref{tab:4fgamma} and~\ref{tab:6fgamma} show the reduction in
computational complexity for some important processes at a
$e^+e^-$-linear collider including radiative corrections.  Using the
asymmetric keystones can reduce the number of vertices by some~10
to~20 percent relativ to the quoted numbers for symmetric keystones.

%%%%%%%%%%%%%%%%%%%%%%%%%%%%%%%%%%%%%%%%%%%%%%%%%%%%%%%%%%%%%%%%%%%%%%%%
\subsection{Comparisons}
\label{sec:comparisons}

HELAC's~\cite{HELAC:2000} diagnostics report more vertices than O'Mega
for identical amplitudes.  This ranges from comparable numbers for
Standard Model processes with many different flavors to an increase by
50 percent for processes with many identical flavors.  Empirically,
O'Mega's straight line code is twice as fast as HELAC's DO-loops for
identical optimizing Fortran95 compilers (not counting HELAC's
initialization phase).  Together this results in an improved
performance by a factor of two to three.

The numerical efficiency of O'Mega's Fortran95 runtime library is
empirically identical to HELAS~\cite{HELAS}. Therefore, O'Mega's
performance can directly be compared to
MADGRAPH's~\cite{MADGRAPH:1994} by comparing the number of vertices.
For $2\to5$-processes in the Standard Model, O'Mega's advantage in
performance is about a factor of two and grows from there.

The results have been compared with MADGRAPH~\cite{MADGRAPH:1994} for
many Standard Model processes and numerical agreement at the level
of~$10^{-11}$ has been found with double precision floating point
arithmetic.

%%%%%%%%%%%%%%%%%%%%%%%%%%%%%%%%%%%%%%%%%%%%%%%%%%%%%%%%%%%%%%%%%%%%%%%%
\subsection{Applications}
O'Mega generated amplitudes are used in the omnipurpose
event generator generator WHIZARD~\cite{Kilian:WHIZARD}.  The first
complete experimental study of vector boson scattering in six fermion
production for linear collider
physics~\cite{Chierici/Kobel/Rosati:2000:TDR-backup} has been
facilitated by O'Mega and WHIZARD.

%%%%%%%%%%%%%%%%%%%%%%%%%%%%%%%%%%%%%%%%%%%%%%%%%%%%%%%%%%%%%%%%%%%%%%%%
\section*{Acknowledgements}
We thank to Wolfgang Kilian for providing the WHIZARD environment that
turns our numbers into real events with unit weight.  Thanks to the
ECFA/DESY workshops and their participants for providing a showcase.
This research is supported by Bundesministerium f\"ur Bildung und
Forschung, Germany, (05\,HT9RDA) and Deutsche Forschungsgemeinschaft
(MA\,676/6-1).

Finally, thanks to the Caml and Objective Caml teams at INRIA for the
lean and mean implementation of a programming language that does not
insult the programmer's intelligence.

%%%%%%%%%%%%%%%%%%%%%%%%%%%%%%%%%%%%%%%%%%%%%%%%%%%%%%%%%%%%%%%%%%%%%%%%

%%%%%%%%%%%%%%%%%%%%%%%%%%%%%%%%%%%%%%%%%%%%%%%%%%%%%%%%%%%%%%%%%%%%%%%%
\appendix
\section{Installing O'Mega}
\label{sec:installation}
\aname{installation}{}%
\subsection{Sources}
O'Mega is Free Software and the sources can be obtained from
%%% \ahrefurl{http://heplix.ikp.physik.tu-darmstadt.de/\home{ohl}/omega/} or from
\url{http://heplix.ikp.physik.tu-darmstadt.de/~ohl/omega/} or from
\ahrefurl{ftp://heplix.ikp.physik.tu-darmstadt.de/pub/ohl/omega}.
The command
\begin{code}
ohl@thopad:~mc$ zcat omega-current.tar.gz | tar xf -
\end{code}
will unpack the sources to the directory \url{omega}.  The
subdirectories of \url{omega} are
\begin{files}
  \item[bin] contains executable instances of O'Mega: \url{f90_SM.bin}
    (\url{f90_SM.opt} if the sytem is supported by O'Caml's native
    code compiler), \url{f90_QED.bin}, etc.
  \item[doc] contains \LaTeX{} sources of user documentation.
  \item[examples] contains currently no supported examples.
  \item[lib] contains library support for targets (Fortran90/95 modules, etc.).
  \item[src] contains the unabridged and uncensored sources of O'Mega,
    including comments.
  \item[tests] contains a battery of regression tests.  Most tests
    require Madgraph~\cite{MADGRAPH:1994}.
  \item[web] contains the `woven' sources, i.\,e.~a pretty printed
    version of the source including \LaTeX{} documentation.  Weaving
    the sources requires programs, \url{ocamlweb} and \url{noweb}.
    A complete PostScript file is available from the same place as
    the O'Mega sources. (It is not required for the end user to read this.)
\end{files}
%%%%%%%%%%%%%%%%%%%%%%%%%%%%%%%%%%%%%%%%%%%%%%%%%%%%%%%%%%%%%%%%%%%%%%%%
\subsection{Prerequisites}
\subsubsection{Objective Caml (a.\,k.\,a.~O'Caml)}
You need version 3.00 or higher\footnote{O'Mega can probably be made
to work with versions 2.0x.  Given the simplicity of building O'Caml
version 3.00, this is not worth the effort.}.  You can get it
from~\ahrefurl{http://pauillac.inria.fr/ocaml/}.  There are precompiled
binaries for some popular systems and complete sources.  Building from
source is straightforward (just follow the instructions in the
file~\url{INSTALL} in the toplevel directory, the defaults are almost
always sufficient) and takes $\mathcal{O}(10)$ minutes on a modern
desktop system.  If available for your system (cf.~the file
\url{README} in the toplevel directory), you should build the native
code compiler.
%%%%%%%%%%%%%%%%%%%%%%%%%%%%%%%%%%%%%%%%%%%%%%%%%%%%%%%%%%%%%%%%%%%%%%%%
\subsubsection{GNU \texttt{make}}
This should be available for any system of practical importance and it
makes no sense to waste physicist's time on supporting all
incompatible flavors of \url{make} in existence.  GNU \url{make} is
the default on Linux systems and is often available as \url{gmake} on
commercial Unices.
%%%%%%%%%%%%%%%%%%%%%%%%%%%%%%%%%%%%%%%%%%%%%%%%%%%%%%%%%%%%%%%%%%%%%%%%
\subsubsection{Fortran90/95 Compiler}
Not required for compiling or running O'Mega, but Fortran90/95 is
currently the only fully supported target.
%%%%%%%%%%%%%%%%%%%%%%%%%%%%%%%%%%%%%%%%%%%%%%%%%%%%%%%%%%%%%%%%%%%%%%%%
\subsection{Configuration}
Before the next step, O'Caml must have been installed.  Configuration
is performed automatically by testing some system features with the
command 
\begin{code}
$ ./configure
\end{code}
See
\begin{code}
$ ./configure --help
\end{code}
for additional options.  NB: The use of the options
\url{--enable-gui} and \url{--enable-unsupported} is strongly
discouraged.  The resulting programs require additional prerequisites
and even if you can get them to compile, the results are unpredictable
and we will not answer any questions about them.  NB: \url{configure}
keeps it's state in \url{config.cache}.  If you want to reconfigure
after adding new libraries to your system, you should remove
\url{config.cache} before running \url{configure}.
%%%%%%%%%%%%%%%%%%%%%%%%%%%%%%%%%%%%%%%%%%%%%%%%%%%%%%%%%%%%%%%%%%%%%%%%
\subsection{Compilation}
The command
\begin{code}
$ make bin
\end{code}
will build the byte code executables.  For each pairing of physics
model and target language, there will be one executable.
\begin{code}
$ make opt
\end{code}
will build the native code executables if the sytem is supported by
O'Caml's native code compiler and it is installed.  The command
\begin{code}
$ make f95
\end{code}
will build the Fortran90/95 library and requires, obviously, a
Fortran90/95 compiler.
%%%%%%%%%%%%%%%%%%%%%%%%%%%%%%%%%%%%%%%%%%%%%%%%%%%%%%%%%%%%%%%%%%%%%%%%
\section{Running O'Mega}
\label{sec:running}
O'Mega is a simple application that takes parameters from the
commandline and writes results to the standard output
device\footnote{In the future, other targets than Fortran90/95 might
require more than one output file (e.\,g.~source files and header
files for \texttt{C}/\texttt{C++}).  In this case the filenames will be
specified by commandline parameters.}
(diagnostics go to the standard error device).  E.\,g., the UNIX
commandline
\begin{code}
$ ./bin/f90_SM.opt e+ e- e+ nue ubar d > cc20_amplitude.f95
\end{code}
will cause O'Mega to write a Fortran95 module containing the Standard
Model tree level scattering amplitude for~$e^+e^-\to e^+\nu_e\bar{u}d$
to the file \url{cc20_amplitude.f95}.  Particles can be combined with
colons.  E.\,g.,
\begin{code}
$ ./bin/f90_SM.opt ubar:u:dbar:d ubar:u:dbar:d e+:mu+ e-:mu- > dy.f95
\end{code}
will cause O'Mega to write a Fortran95 module containing the Standard
Model tree level parton scattering amplitudes for all Drell-Yan
processes to the file \url{dy.f95}.\par
A synopsis of the available options, in particular the particle names,
can be requested by giving an illegal option, e.\,g.:
\begin{code}
$ ./bin/f90_SM.opt -?
./bin/f90_SM.opt: unknown option `-?'.
usage: ./bin/f90_SM.opt [options] [e-|nue|u|d|e+|nuebar|ubar|dbar\
   |mu-|numu|c|s|mu+|numubar|cbar|sbar|tau-|nutau|t|b\
   |tau+|nutaubar|tbar|bbar|A|Z|W+|W-|g|H|phi+|phi-|phi0]
  -target:function function name
  -target:90 don't use Fortran95 features that are not in Fortran90
  -target:kind real and complex kind (default: omega_prec)
  -target:width approx. line length
  -target:module module name
  -target:use use module
  -target:whizard include WHIZARD interface
  -model:constant_width use constant width (also in t-channel)
  -model:fudged_width use fudge factor for charge particle width
  -model:custom_width use custom width
  -model:cancel_widths use vanishing width
  -warning: check arguments and print warning on error
  -error: check arguments and terminate on error
  -warning:a check # of input arguments and print warning on error
  -error:a check # of input arguments and terminate on error
  -warning:h check input helicities and print warning on error
  -error:h check input helicities and terminate on error
  -warning:m check input momenta and print warning on error
  -error:m check input momenta and terminate on error
  -warning:g check internal Ward identities and print warning on error
  -error:g check internal Ward identities and terminate on error
  -forest ???
  -revision print revision control information
  -quiet don't print a summary
  -summary print only a summary
  -params print the model parameters
  -poles print the Monte Carlo poles
  -dag print minimal DAG
  -full_dag print complete DAG
  -file read commands from file 
\end{code}
%%%%%%%%%%%%%%%%%%%%%%%%%%%%%%%%%%%%%%%%%%%%%%%%%%%%%%%%%%%%%%%%%%%%%%%%
\subsection{General Options}
\begin{options}
  \item[-warning:] include code that checks the supplied arguments and
    prints a warning in case of an error.
  \item[-warning:a] check the number of input arguments (momenta and
    spins) and print a warning in case of an error.
  \item[-warning:h] check the values of the input helicities
    and print a warning in case of an error.
  \item[-warning:m] check the values of the input momenta
    and print a warning in case of an error.
  \item[-warning:g] check internal Ward identities
    and print a warning in case of an error (not supported yet!).
  \item[-error:] like \verb+-warning:+ but terminates on error.
  \item[-error:a] like \verb+-warning:a+ but terminates on error.
  \item[-error:h] like \verb+-warning:h+ but terminates on error.
  \item[-error:m] like \verb+-warning:m+ but terminates on error.
  \item[-error:g] like \verb+-warning:g+ but terminates on error.
  %item[-forest] ???
  \item[-revision] print revision control information
  \item[-quiet] don't print a summary
  \item[-summary] print only a summary
  \item[-params] print the model parameters
  \item[-poles] print the Monte Carlo poles in a format understood by
    the WHIZARD program~\cite{Kilian:WHIZARD}.
  \item[-dag] print the reduced DAG in a format understood by the
    \texttt{dot} program.
  \item[-full\_dag] print the complete DAG in a format understood by the
    \texttt{dot} program.
  \item[-file] read commands from file 
\end{options}
%%%%%%%%%%%%%%%%%%%%%%%%%%%%%%%%%%%%%%%%%%%%%%%%%%%%%%%%%%%%%%%%%%%%%%%%
\subsection{Model Options}
\subsubsection{Standard Model}
\begin{options}
  \item[-model:constant\_width] use constant width (also in  $t$-channel)
  \item[-model:fudged\_width] use fudge factor for charge particle width
  \item[-model:custom\_width] use custom width
  \item[-model:cancel\_widths] use vanishing width
\end{options}
%%%%%%%%%%%%%%%%%%%%%%%%%%%%%%%%%%%%%%%%%%%%%%%%%%%%%%%%%%%%%%%%%%%%%%%%
\subsection{Target Options}
\subsubsection{Fortran90/95}
\begin{options}
  \item[-target:function] function name
  \item[-target:90] don't use Fortran95 features that are not in Fortran90
  \item[-target:kind] real and complex kind (default: \verb+omega_prec+)
  \item[-target:width] approx. line length
  \item[-target:module] module name
  \item[-target:use] use module
  \item[-target:whizard] include WHIZARD interface
\end{options}
%%%%%%%%%%%%%%%%%%%%%%%%%%%%%%%%%%%%%%%%%%%%%%%%%%%%%%%%%%%%%%%%%%%%%%%%
\section{Using O'Mega's Output}
\label{sec:using}
The structure of the outputfile, the calling convention and the
required libraries depends on the target language, of course.
%%%%%%%%%%%%%%%%%%%%%%%%%%%%%%%%%%%%%%%%%%%%%%%%%%%%%%%%%%%%%%%%%%%%%%%%
\subsection{Fortran90/95}
The Fortran95 module written by O'Mega has the following signature
\begin{code}
module omega_amplitude
\end{code}
%%%%%%%%%%%%%%%%%%%%%%%%%%%%%%%%%%%%%%%%%%%%%%%%%%%%%%%%%%%%%%%%%%%%%%%%
\subsubsection{Libraries}
The imported Fortran modules are
\begin{files}
  \item[omega\_kinds] defines \verb+omega_prec+, which can be whatever
    the Fortran compiler supports.  NB: the support libraries have not
    yet been tuned to give reliable answers for amplitudes with gauge
    cancellations in single precision.
  \item[omega95] defines the vertices for Dirac spinors in the chiral
    representation and vectors.
  \item[omega95\_bispinors] is an alternative that defines the
    vertices for Dirac and Majorana spinors in the chiral
    representation and vectors using the Feynman rules
    of~\cite{Denner/etal:Majorana}.
  \item[omega\_parameters] defines the coupling constants
\end{files}
\begin{code}
  use omega_kinds
  use omega95
  use omega_parameters
  implicit none
  private
\end{code}
%%%%%%%%%%%%%%%%%%%%%%%%%%%%%%%%%%%%%%%%%%%%%%%%%%%%%%%%%%%%%%%%%%%%%%%%
\subsubsection{Summary of Exported Functions}
The functions and subroutines experted by the Fortran95 module are
\begin{itemize}
  \item the scattering amplitude in different flavor bases (arrays of
    PDG codes or internal numbering):
\begin{code}
  public :: amplitude, amplitude_f, amplitude_1, amplitude_2
\end{code}
  \item the scattering amplitude with heuristics supressing vanishing
   helicity combinations:
\begin{code}
  public :: amplitude_nonzero, amplitude_f_nonzero, &
    amplitude_1_nonzero, amplitude_2_nonzero
\end{code}
  \item the squared scattering amplitude summed over helicity states
\begin{code}
  public :: spin_sum_sqme, spin_sum_sqme_1, sum_sqme
  public :: spin_sum_sqme_nonzero, spin_sum_sqme_1_nonzero, &
    sum_sqme_nonzero
\end{code}
  \item ``scattering'' a general density matrix
\begin{code}
  public :: scatter, scatter_nonzero
\end{code}
  \item ``scattering'' a diagonal density matrix
\begin{code}
  public :: scatter_diagonal, scatter_diagonal_nonzero
\end{code}
  \item inquiry and maintenance functions
\begin{code}
  public :: allocate_zero
  public :: multiplicities, multiplicities_in, multiplicities_out
  public :: number_particles, &
    number_particles_in, number_particles_out
  public :: number_spin_states, &
    number_spin_states_in, number_spin_states_out, &
    spin_states, spin_states_in, spin_states_out
  public :: number_flavor_states, &
    number_flavor_states_in, number_flavor_states_out, &
    flavor_states, flavor_states_in, flavor_states_out
  public :: number_flavor_zeros, &
    number_flavor_zeros_in, number_flavor_zeros_out, &
    flavor_zeros, flavor_zeros_in, flavor_zeros_out
  public :: create, reset, destroy
\end{code}
\end{itemize}
%%%%%%%%%%%%%%%%%%%%%%%%%%%%%%%%%%%%%%%%%%%%%%%%%%%%%%%%%%%%%%%%%%%%%%%%
\subsubsection{Maintenance Functions}
They currently do nothing, but are here for
WHIZARD's~\cite{Kilian:WHIZARD} convenience
\begin{files}
  \item[\texttt{create}] is called only once at the very beginning.
  \item[\texttt{reset}] is called whenever parameters are changed.
  \item[\texttt{destroy}] is called at most once at the very end.
\end{files}
\begin{code}
  subroutine create ()
  end subroutine create
  subroutine reset ()
  end subroutine reset
  subroutine destroy ()
  end subroutine destroy
\end{code}
\aname{specific/allocate}{}%
Allocate an array of the size used by the heuristic that suppresses
vanishing helicity combinations
\begin{code}
  interface allocate_zero
     module procedure allocate_zero_1, allocate_zero_2
  end interface
\end{code}
for join numbering of in and out states
\begin{code}
  subroutine allocate_zero_1 (zero)
    integer, dimension(:,:), pointer :: zero
  end subroutine allocate_zero_index
\end{code}
and for separate numbering of in and out states
\begin{code}
  subroutine allocate_zero_2 (zero)
    integer, dimension(:,:,:,:), pointer :: zero
  end subroutine allocate_zero_index_inout
\end{code}
%%%%%%%%%%%%%%%%%%%%%%%%%%%%%%%%%%%%%%%%%%%%%%%%%%%%%%%%%%%%%%%%%%%%%%%%
\subsubsection{Inquiry Functions}
\aname{specific/numbers/states}{}%
The total number of particles, the number of incoming particles and
the number of outgoing particles:
\begin{code}
  pure function number_particles () result (n)
    integer :: n
  end function number_particles
  pure function number_particles_in () result (n)
    integer :: n
  end function number_particles_in
  pure function number_particles_out () result (n)
    integer :: n
  end function number_particles_out
\end{code}
The spin states of all particles that can give non-zero results and
their number.  The tables are interpreted as
\begin{files}
  \item[\texttt{s(1:,i)}] contains the helicities for each particle
    for the \verb+i+th helicity combination.
\end{files}
\begin{code}
  pure function number_spin_states () result (n)
    integer :: n
  end function number_spin_states
  pure subroutine spin_states (s)
    integer, dimension(:,:), intent(inout) :: s
  end subroutine spin_states
\end{code}
The spin states of the incoming particles that can give non-zero
results and their number:
\begin{code}
  pure function number_spin_states_in () result (n)
    integer :: n
  end function number_spin_states_in
  pure subroutine spin_states_in (s)
    integer, dimension(:,:), intent(inout) :: s
  end subroutine spin_states_in
\end{code}
The spin states of the outgoing particles that can give non-zero
results and their number:
\begin{code}
  pure function number_spin_states_out () result (n)
    integer :: n
  end function number_spin_states_out
  pure subroutine spin_states_out (s)
    integer, dimension(:,:), intent(inout) :: s
  end subroutine spin_states_out
\end{code}
The flavor combinations of all particles that can give non-zero
results and their number.  The tables are interpreted as
\begin{files}
  \item[\texttt{f(1:,i)}] contains the PDG particle code for each
    particle for the \verb+i+th helicity combination.
\end{files}
\begin{code}
  pure function number_flavor_states () result (n)
    integer :: n
  end function number_flavor_states
  pure subroutine flavor_states (f)
    integer, dimension(:,:), intent(inout) :: f
  end subroutine flavor_states
\end{code}
The flavor combinations of the incoming particles that can give
non-zero results and their number.
\begin{code}
  pure function number_flavor_states_in () result (n)
    integer :: n
  end function number_flavor_states_in
  pure subroutine flavor_states_in (f)
    integer, dimension(:,:), intent(inout) :: f
  end subroutine flavor_states_in
\end{code}
The flavor combinations of the outgoing particles that can give
non-zero results and their number.
\begin{code}
  pure function number_flavor_states_out () result (n)
    integer :: n
  end function number_flavor_states_out
  pure subroutine flavor_states_out (f)
    integer, dimension(:,:), intent(inout) :: f
  end subroutine flavor_states_out
\end{code}
The flavor combinations of all particles that always can give
a zero result and their number:
\begin{code}
  pure function number_flavor_zeros () result (n)
    integer :: n
  end function number_flavor_zeros
  pure subroutine flavor_zeros (f)
    integer, dimension(:,:), intent(inout) :: f
  end subroutine flavor_zeros
\end{code}
The flavor combinations of the incoming particles that always can give
a zero result and their number:
\begin{code}
  pure function number_flavor_zeros_in () result (n)
    integer :: n
  end function number_flavor_zeros_in
  pure subroutine flavor_zeros_in (f)
    integer, dimension(:,:), intent(inout) :: f
  end subroutine flavor_zeros_in
\end{code}
The flavor combinations of the outgoing particles that always can give
a zero result and their number:
\begin{code}
  pure function number_flavor_zeros_out () result (n)
    integer :: n
  end function number_flavor_zeros_out
  pure subroutine flavor_zeros_out (f)
    integer, dimension(:,:), intent(inout) :: f
  end subroutine flavor_zeros_out
\end{code}
\aname{specific/multiplicities}{}%
The same initial and final state can appear more than once in the
tensor product and we must avoid double counting.
\begin{code}
  pure subroutine multiplicities (a)
    integer, dimension(:), intent(inout) :: a
  end subroutine multiplicities
\end{code}
\begin{code}
  pure subroutine multiplicities_in (a)
    integer, dimension(:), intent(inout) :: a
  end subroutine multiplicities_in
\end{code}
\begin{code}
  pure subroutine multiplicities_out (a)
    integer, dimension(:), intent(inout) :: a
  end subroutine multiplicities_out
\end{code}
%%%%%%%%%%%%%%%%%%%%%%%%%%%%%%%%%%%%%%%%%%%%%%%%%%%%%%%%%%%%%%%%%%%%%%%%
\subsubsection{Amplitude}
\aname{specific/amplitude}{}%
The function arguments of of the amplitude are
\begin{files}
  \item[\texttt{k(0:3,1:)}] are the particle momenta: \verb+k(0:3,1)+ and
    \verb+k(0:3,2)+ are the incoming momenta, \verb+k(0:3,3:)+ are the
    outgoing momenta.  \emph{All} momenta are the physical momenta,
    i.\,e.~forward time-like or light-like.  The signs of the incoming
    momenta are flipped \emph{internally}.  Unless asked by a commandline
    parameter, O'Mega will not check the validity of the momenta.
  \item[\texttt{s(1:)}] are the helicities in the same order as the
    momenta.  $s=\pm1$ signify $s=\pm1/2$ for fermions.  $s=0$ makes no
    sense for fermions and massless vector bosons
    $s=4$ signifies an unphysical polarization for vector boson
    that the users are \emph{not} supposed to use.  Unless asked by a
    commandline parameter, O'Mega will not check the validity of the
    helicities.
  \item[\texttt{f(1:)}] are the PDG particle codes in the same order as the
    momenta.
\end{files}
\begin{code}
  pure function amplitude (k, s, f) result (amp)
    real(kind=omega_prec), dimension(0:,:), intent(in) :: k
    integer, dimension(:), intent(in) :: s, f
    complex(kind=omega_prec) :: amp
  end function amplitude
\end{code}
Identical to \verb+amplitude (k, s, flavors(:,f))+, where
\verb+flavors+ has been filled by \verb+flavor_states+:
\begin{code}
  pure function amplitude_f (k, s, f) result (amp)
    real(kind=omega_prec), dimension(0:,:), intent(in) :: k
    integer, dimension(:), intent(in) :: s
    integer, intent(in) :: f
    complex(kind=omega_prec) :: amp
  end function amplitude_f
\end{code}
Identical to \verb+amplitude (k, spins(:,s), flavors(:,f))+, where
\verb+spins+ has been filled by \verb+spin_states+ and
\verb+flavors+ has been filled by \verb+flavor_states+:
\begin{code}
  pure function amplitude_1 (k, s, f) result (amp)
    real(kind=omega_prec), dimension(0:,:), intent(in) :: k
    integer, intent(in) :: s, f
    complex(kind=omega_prec) :: amp
  end function amplitude_1
\end{code}
Similar to \verb+amplitude_1+, but with separate incoming and
outgoing particles:
\begin{code}
  pure function amplitude_2 &
       (k, s_in, f_in, s_out, f_out) result (amp)
    real(kind=omega_prec), dimension(0:,:), intent(in) :: k
    integer, intent(in) :: s_in, f_in, s_out, f_out
    complex(kind=omega_prec) :: amp
  end function amplitude_2
\end{code}
\aname{specific/amplitude/nonzero}{}%
The following are subroutines and not functions, since Fortran95
restricts arguments of pure functions to \verb+intent(in)+, but we
need to update the counter for vanishing amplitudes.
\begin{files}
  \item[\texttt{zero(1:,1:)}] an array containing the number of times
    a combination of spin index and flavor index yielded a vanishing
    amplitude.  After a certain threshold, these combinations will be
    skipped. \verb+allocate_zero+ will allocate the correct size.
  \item[\texttt{n}] the current event count
\end{files}
\begin{code}
  pure subroutine amplitude_nonzero (amp, k, s, f, zero, n)
    complex(kind=omega_prec), intent(out) :: amp
    real(kind=omega_prec), dimension(0:,:), intent(in) :: k
    integer, dimension(:), intent(in) :: s, f
    integer, dimension(:,:), intent(inout) :: zero
    integer, intent(in) :: n
  end subroutine amplitude_nonzero
\end{code}
\begin{code}
  pure subroutine amplitude_1_nonzero (amp, k, s, f, zero, n)
    complex(kind=omega_prec), intent(out) :: amp
    real(kind=omega_prec), dimension(0:,:), intent(in) :: k
    integer, intent(in) :: s, f
    integer, dimension(:,:), intent(inout) :: zero
    integer, intent(in) :: n
  end subroutine amplitude_1_nonzero
\end{code}
\begin{code}
  pure subroutine amplitude_f_nonzero &
         (amp, k, s, f, zero, n)
    complex(kind=omega_prec), intent(out) :: amp
    real(kind=omega_prec), dimension(0:,:), intent(in) :: k
    integer, dimension(:), intent(in) :: s
    integer, intent(in) :: f
    integer, dimension(:,:), intent(inout) :: zero
    integer, intent(in) :: n
  end subroutine amplitude_f_nonzero
\end{code}
\begin{files}
  \item[\texttt{zero(1:,1:,1:,1:)}] an array containing the number of
    times a combination of incoming and outgoing spin indices and
    flavor indices yielded a vanishing amplitude.
    \verb+allocate_zero+ will allocate the correct size.
\end{files}
\begin{code}
  pure subroutine amplitude_2_nonzero &
       (amp, k, s_in, f_in, s_out, f_out, zero, n)
    complex(kind=omega_prec), intent(out) :: amp
    real(kind=omega_prec), dimension(0:,:), intent(in) :: k
    integer, intent(in) :: s_in, f_in, s_out, f_out
    integer, dimension(:,:,:,:), intent(inout) :: zero
    integer, intent(in) :: n
  end subroutine amplitude_2_nonzero
\end{code}
%%%%%%%%%%%%%%%%%%%%%%%%%%%%%%%%%%%%%%%%%%%%%%%%%%%%%%%%%%%%%%%%%%%%%%%%
\subsubsection{Summation}
\aname{specific/sum}{}%
The the sums of squared matrix elements, the optional mask \url{smask}
can be used to sum only a subset of helicities or flavors.
\begin{code}
  pure function spin_sum_sqme (k, f, smask) result (amp2)
    real(kind=omega_prec), dimension(0:,:), intent(in) :: k
    integer, dimension(:), intent(in) :: f
    logical, dimension(:), intent(in), optional :: smask
    real(kind=omega_prec) :: amp2
  end function spin_sum_sqme
\end{code}
\begin{code}
  pure function spin_sum_sqme_1 (k, f, smask) result (amp2)
    real(kind=omega_prec), dimension(0:,:), intent(in) :: k
    integer, intent(in) :: f
    logical, dimension(:), intent(in), optional :: smask
    real(kind=omega_prec) :: amp2
  end function spin_sum_sqme_1
\end{code}
\begin{code}
  pure function sum_sqme (k, smask, fmask) result (amp2)
    real(kind=omega_prec), dimension(0:,:), intent(in) :: k
    logical, dimension(:), intent(in), optional :: smask, fmask
    real(kind=omega_prec) :: amp2
  end function sum_sqme
\end{code}
\aname{specific/sum/nonzero}{}%
\begin{code}
  pure subroutine spin_sum_sqme_nonzero (amp2, k, f, zero, n, smask)
    real(kind=omega_prec), intent(out) :: amp2
    real(kind=omega_prec), dimension(0:,:), intent(in) :: k
    integer, dimension(:), intent(in) :: f
    integer, dimension(:,:), intent(inout) :: zero
    integer, intent(in) :: n
    logical, dimension(:), intent(in), optional :: smask
  end subroutine spin_sum_sqme_nonzero
\end{code}
\begin{code}
  pure subroutine spin_sum_sqme_1_nonzero (amp2, k, f, zero, n, smask)
    real(kind=omega_prec), intent(out) :: amp2
    real(kind=omega_prec), dimension(0:,:), intent(in) :: k
    integer, intent(in) :: f
    integer, dimension(:,:), intent(inout) :: zero
    integer, intent(in) :: n
    logical, dimension(:), intent(in), optional :: smask
  end subroutine spin_sum_sqme_1_nonzero
\end{code}
\begin{code}
  pure subroutine sum_sqme_nonzero (amp2, k, zero, n, smask, fmask)
    real(kind=omega_prec), intent(out) :: amp2
    real(kind=omega_prec), dimension(0:,:), intent(in) :: k
    integer, dimension(:,:), intent(inout) :: zero
    integer, intent(in) :: n
    logical, dimension(:), intent(in), optional :: smask, fmask
  end subroutine sum_sqme_masked_nonzero
\end{code}
%%%%%%%%%%%%%%%%%%%%%%%%%%%%%%%%%%%%%%%%%%%%%%%%%%%%%%%%%%%%%%%%%%%%%%%%
\subsubsection{Density Matrix Transforms}
\aname{specific/scatter}{}%
There are also utility functions that implement the transformation of
density matrices directly
\begin{equation}
  \rho \to \rho' = T \rho T^{\dagger}
\end{equation}
i.\,e.
\begin{equation}
  \rho'_{ff'} = \sum_{ii'} T_{fi} \rho_{ii'} T^{*}_{f'i'}
\end{equation}
and avoid double counting
\begin{code}
  pure subroutine scatter_correlated (k, rho_in, rho_out)
    real(kind=omega_prec), dimension(0:,:), intent(in) :: k
    complex(kind=omega_prec), dimension(:,:,:,:), &
      intent(in) :: rho_in
    complex(kind=omega_prec), dimension(:,:,:,:), &
      intent(inout) :: rho_out
  end subroutine scatter_correlated
\end{code}
\begin{code}
  pure subroutine scatter_correlated_nonzero &
         (k, rho_in, rho_out, zero, n)
    real(kind=omega_prec), dimension(0:,:), intent(in) :: k
    complex(kind=omega_prec), dimension(:,:,:,:), &
      intent(in) :: rho_in
    complex(kind=omega_prec), dimension(:,:,:,:), &
      intent(inout) :: rho_out
    integer, dimension(:,:,:,:), intent(inout) :: zero
    integer, intent(in) :: n
  end subroutine scatter_correlated_nonzero
\end{code}
In no off-diagonal density matrix elements of the initial state are
known, the computation can be performed more efficiently:
\begin{equation}
  \rho'_{f} = \sum_i T_{fi} \rho_{i} T^{*}_{fi}
            = \sum_i |T_{fi}|^2 \rho_{i}
\end{equation}
\begin{code}
  pure subroutine scatter_diagonal (k, rho_in, rho_out)
    real(kind=omega_prec), dimension(0:,:), intent(in) :: k
    real(kind=omega_prec), dimension(:,:), intent(in) :: rho_in
    real(kind=omega_prec), dimension(:,:), intent(inout) :: rho_out
  end subroutine scatter_diagonal
\end{code}
\begin{code}
  pure subroutine scatter_diagonal_nonzero &
         (k, rho_in, rho_out, zero, n)
    real(kind=omega_prec), dimension(0:,:), intent(in) :: k
    real(kind=omega_prec), dimension(:,:), intent(in) :: rho_in
    real(kind=omega_prec), dimension(:,:), intent(inout) :: rho_out
    integer, dimension(:,:,:,:), intent(inout) :: zero
    integer, intent(in) :: n
  end subroutine scatter_diagonal_nonzero
\end{code}
Finis.
\begin{code}
end module omega_amplitude
\end{code}
NB: the name of the module can be changed by a
commandline parameter and Fortran95 features like \verb+pure+ can be
disabled as well.
%%%%%%%%%%%%%%%%%%%%%%%%%%%%%%%%%%%%%%%%%%%%%%%%%%%%%%%%%%%%%%%%%%%%%%%%
\subsection{FORTRAN77}
The preparation of a FORTRAN77 target is straightforward, but tedious
and will only be considered if there is sufficient demand and support.
%%%%%%%%%%%%%%%%%%%%%%%%%%%%%%%%%%%%%%%%%%%%%%%%%%%%%%%%%%%%%%%%%%%%%%%%
\subsection{HELAS}
This target for the HELAS library~\cite{HELAS} is incomplete and no
longer maintained.  It was used as an early benchmark for the
Fortran90/95 library.  No vector boson selfcouplings are supported.
%%%%%%%%%%%%%%%%%%%%%%%%%%%%%%%%%%%%%%%%%%%%%%%%%%%%%%%%%%%%%%%%%%%%%%%%
\subsection{\texttt{C}, \texttt{C++} \&\ Java}
These targets does not exist yet and we solicit suggestions from
\texttt{C++} and Java experts on useful calling conventions and
suppport libraries that blend well with the HEP environments based on
these languages.  At least one of the authors believes that Java would
be a better choice, but the political momentum behind \texttt{C++}
might cause an early support for \texttt{C++} anyway.
%%%%%%%%%%%%%%%%%%%%%%%%%%%%%%%%%%%%%%%%%%%%%%%%%%%%%%%%%%%%%%%%%%%%%%%%
\section{Extending O'Mega}
\label{sec:extensions}
\subsection{Adding A New Physics Model}
Currently, this still requires to write O'Caml code.  This is not as
hard as it might sound, because an inspection of \url{bin/models.ml}
shows that all that is required are some tables of Feynman rules that
can easily be written by copying and modifyng an existing example,
after consulting with \url{src/couplings.mli} or the corresponding
chapter in the woven source.
In fact, having the full power of O'Caml at one's disposal is
very helpful for avoiding needless repetition.

Nevertheless, in the near future, there will be some special models
that can read model specifications from external files.  The first one
of its kind will read CompHEP~\cite{CompHEP} model files.  Later there
will be a native O'Mega model file format, but it will probably go
through some iterations.
%%%%%%%%%%%%%%%%%%%%%%%%%%%%%%%%%%%%%%%%%%%%%%%%%%%%%%%%%%%%%%%%%%%%%%%%
\subsection{Adding A New Target Language}
This will always require to write O'Caml code, which is again not too
hard.  In addition a library for vertices will be required, unless the
target performs complete inlining.  NB: an early experiment with
inlining Fortran proved to be an almost complete failure on Linux/Intel PCs.
The inlined code was huge, absolutely unreadable and only marginally
faster.  The bulk of the computational cost is always in the vertex
evaluations and function calls create in comparison negligible costs.
This observation is system dependent, of course, and inlining
might be beneficial for other architectures with better floating point
performance, after all.
%%%%%%%%%%%%%%%%%%%%%%%%%%%%%%%%%%%%%%%%%%%%%%%%%%%%%%%%%%%%%%%%%%%%%%%%
%BEGIN IMAGE
\end{empfile}
\end{fmffile}
%END IMAGE
\end{document}